\documentstyle[11pt]{article}
\begin{document}
\setlength{\baselineskip}{15pt}
\title{Characterization and solvability of \\ quasipolynomial symplectic mappings}
\author{ Benito Hern\'{a}ndez--Bermejo$^{\:(1)}$  \and L\'{e}on Brenig$^{\:(2)}$}
\date{}

\maketitle
\begin{center}
{\em \mbox{} \\
Service de Physique Th\'{e}orique et Math\'{e}matique. Universit\'{e} Libre de Bruxelles. \\
Campus Plaine -- CP 231. Boulevard du Triomphe, B-1050 Brussels, Belgium.}
\end{center}

\mbox{}

\mbox{}

\begin{abstract}
Quasipolynomial (or QP) mappings  constitute a wide generalization of the well-known Lotka-Volterra mappings, of importance in different fields such as population dynamics, Physics, Chemistry or Economy. In addition, QP mappings are a natural discrete-time analog of the continuous QP systems, which have been extensively used in different pure and applied domains. After presenting the basic definitions and properties of QP mappings in a previous article \cite{bl1}, the purpose of this work is to focus on their characterization by considering the existence of symplectic QP mappings. In what follows such QP symplectic maps are completely characterized. Moreover, use of the QP formalism can be made in order to demonstrate that all QP symplectic mappings have an analytical solution that is explicitly and generally constructed. Examples are given.
\end{abstract}

\mbox{}

\mbox{}

\mbox{}



\mbox{}

Short title: Symplectic QP mappings.

\mbox{}

\mbox{}

\mbox{}

\mbox{}

\mbox{}

\mbox{}

\mbox{}

\mbox{}

\mbox{}

\noindent $^{(1)}$ {\bf Present address:} E.S.C.E.T. (Edificio Departamental II). Universidad Rey Juan Carlos. Calle Tulip\'{a}n S/N. 28933--M\'{o}stoles--Madrid (Spain). E-mail: bhernandez@escet.urjc.es

\mbox{}

\noindent $^{(2)}$ {\bf Corresponding author.} Fax: (+ 00 32 2) 650 58 24. E-mail: lbrenig@ulb.ac.be

\pagebreak

\newtheorem{df}{Definition}
\newtheorem{th}{Theorem}
\newtheorem{pr}{Proposition}
\newtheorem{co}{Corollary}

\begin{flushleft}
{\bf 1. Introduction}
\end{flushleft}

In a previous article \cite{bl1} a new family of mappings termed quasipolynomial (QP in what follows) was introduced. In such work it was noted that the interest of QP mappings is twofold: 
\begin{enumerate}
\item They constitute a wide generalization of the well-known Lotka-Volterra mappings, of importance in different fields such as biological modelling \cite{may1}-\cite{lst1}, population dynamics \cite{kr1}-\cite{lw1}, Physics \cite{raj1}-\cite{ur2}, Chemistry \cite{gb1} or Economy \cite{bd1,doh1}. In fact, Lotka-Volterra mappings are not just a particular QP case, but play a central (actually canonical) role in the theory of QP mappings \cite{bl1}.
\item They are a mathematically natural discrete-time analog of the continuous QP 
systems, which have been extensively used in many different mathematical and applied 
contributions (see \cite{bl1} for a bibliography on continuous QP systems and their applications as well as for a detailed analysis of the connection between the discrete and 
the continuous QP formalisms). 
\end{enumerate}
A classification of QP mappings (or even of Lotka-Volterra mappings) is still an open issue. 
However, some preliminary results were already demonstrated in \cite{bl1}. In this sense, an 
important kind of discrete-time systems is the symplectic one \cite{tab,eas}. Symplectic maps are the discrete-time analog of Hamiltonian dynamical systems, and consequently their physical relevance is clear both as models for systems in which there is no dissipation of energy and also as fixed-time maps of Hamiltonian flows. The purpose of this work is to consider the existence of symplectic QP mappings. Actually, in what follows such QP symplectic maps will be completely characterized. Moreover, it will be possible to make use of the QP formalism in order to demonstrate that all QP symplectic maps have an analytical solution which will be explicitly constructed. Additionally, the results obtained uncover a relationship between QP symplectic maps and some families of QP differential systems of interest in the framework of normal form analysis. This perspective will be considered in the last section of the paper.

The structure of the article is the following. In Section 2 some basic facts and properties regarding the QP formalism for mappings are reviewed in order to make the article 
self-contained. Section 3 is devoted to the complete characterization of QP symplectic mappings, while in Section 4 their analytical solutions are constructed. Finally, in Section 5 some concluding remarks are discussed. 

\mbox{}

\mbox{}

\begin{flushleft}
{\bf 2. Overview of the QP formalism for mappings}
\end{flushleft}

The aim of this section is to present an overview of the QP formalism for mappings. The reader is referred to \cite{bl1} for the full details. QP mappings are those of the form
\begin{equation}
\label{qpm}
	x_i(t+1)=x_i(t) \exp \left( \lambda _i + \sum _{j=1}^m A_{ij} \prod _{k=1}^n 
	[x_k(t)]^{B_{jk}} \right)
	\:\: , \:\:\:\:\: i =1, \ldots , n
\end{equation}
where {\em (i)} $m$ is an integer not necessarily equal to $n$; {\em (ii)} index $t$ is an 
integer denoting the discrete time; {\em (iii)} variables $x_i(t)$ are assumed to be positive for $i=1, \ldots ,n$ and for every $t$; and {\em (iv)} $A=(A_{ij})$, $B=(B_{ij})$ and 
$\lambda = ( \lambda _i)$ are real matrices of dimensions $n \times m$, $m \times n$ and $n \times 1$, respectively. Note that this definition implies that matrix $A$ cannot have a column of zeros, and that matrix $B$ cannot have a row of zeros either. The terms 
\[
	\prod _{k=1}^n [x_k(t)]^{B_{jk}} \: , \:\:\: j=1, \ldots ,m
\]
appearing in the exponential of equation (\ref{qpm}) are known as quasimonomials. It is also convenient to introduce an additional matrix, denoted by $M$, which is of dimension $n \times (m+1)$ and is defined as:
\[
	M \equiv ( \lambda \mid A) =
	\left( \begin{array}{cccc}
		\lambda _1 & A_{11} & \ldots  & A_{1m} \\
		\vdots     & \vdots & \mbox{} & \vdots \\
		\lambda _n & A_{n1} & \ldots  & A_{nm} 
	\end{array} \right)
\]
Notice that Lotka-Volterra mappings 
\[
	x_i(p+1)=x_i(p) \exp \left( \lambda _i + \sum _{j=1}^n A_{ij} x_j(p) \right)
	\:\: , \:\:\:\:\: i =1, \ldots , n
\]
are a particular case of QP mapping, namely the one corresponding to $m=n$ and $B$ the $n \times n$ identity matrix.

An important basic property is that the positive orthant is an invariant set for every QP mapping. This is natural in many domains (such as population dynamics) in which the 
system variables are positive by definition. In the QP context, this feature is always present. 

A key set of transformations relating QP mappings are the quasimonomial transformations (QMTs) defined as: 
\[
	x_i(t)= \prod_{j=1}^n [y_j(t)]^{C_{ij}} \:\: , \:\:\: i = 1, \ldots ,n \:\: ; \:\:\: 
	\mid C \mid \neq 0
\]
The form-invariance of QP mappings after a QMT is one of the cornerstones of the formalism. Actually, if we consider a $n$-dimensional QP mapping of matrices $A$, $B$, $\lambda$ (and $M$) and perform a QMT of matrix $C$, the result is another $n$-dimensional QP mapping of matrices $A'$, $B'$, $\lambda '$ (and $M'$) where:
\begin{equation}
\label{mtqmt}
	A' = C^{-1} \cdot A \: , \:\:\: 
	B' = B \cdot C \: , \:\:\: 
	\lambda ' = C^{-1} \cdot \lambda \: , \:\:\: 
	M' = C^{-1} \cdot M 
\end{equation}
Moreover, every QMT relating two QP mappings is a topological conjugacy. Consequently, we not only have a formal invariance between QP systems related by a QMT, but actually a complete dynamical equivalence. These properties imply that the set of all QP mappings related by means of QMTs actually constitute an equivalence class. One important label of such classes is given by the matrix product $B \cdot M$, which is invariant for every equivalence class.

\mbox{}

\mbox{}

\pagebreak
\begin{flushleft}
{\bf 3. Characterization of symplectic QP mappings}
\end{flushleft}

We now focus on the central issue of the article, namely the symplectic property for QP maps. We start by recalling two necessary definitions \cite{eas}:

\begin{df}
\label{df1}
A real $n \times n$ matrix $K$ of even size $n=2s$ and constant entries is said to be symplectic if and only if 
\begin{equation}
\label{sympcond}
	K^T \cdot S \cdot K = S
\end{equation}
where superscript $^T$ means the transpose of a matrix, $S$ is the $n$-dimensional symplectic matrix
\begin{equation}
\label{sympmat}
	S = \left( \begin{array}{rr} 
		O_{s \times s} & -I_{s \times s} \\
		I_{s \times s} &  O_{s \times s} 
	    \end{array} \right)
\end{equation}
and $O$ and $I$ denote the null matrix and identity matrix of the specified sizes, respectively.
\end{df}

It can be demonstrated \cite{eas} that every symplectic matrix has determinant equal to 1. A second necessary definition is:

\begin{df}
\label{df2}
A real mapping on $I \! \! R^n$, with $n=2s$, is said to be symplectic if and only if its Jacobian matrix is a symplectic matrix at each point. 
\end{df}

In our case, these definitions are to be applied in the interior of the positive orthant, according to the definition of QP mappings. We are then in position to state our first main result:

\begin{th}
\label{th1}
A QP mapping (\ref{qpm}) with $n=2s$ is symplectic if and only if the following conditions hold: 
\begin{description}
\item[{\rm a)}] $A_{ij}+A_{s+i,j}=0 \:\:$ for all $i=1, \ldots , s$, and for all $j=1, \ldots , 
m$.

\item[{\rm b)}] $\lambda_{i}+\lambda_{s+i}=0 \:\:$ for all $i=1, \ldots , s$.

\item[{\rm c)}] $A_{ip}B_{pj}=A_{ip}B_{p,s+j}=0 \:\:$ for all $i \neq j $, $\: 1 \leq i,j \leq s$, and for all $p=1, \ldots , m$.

\item[{\rm d)}] $A_{ip}(B_{pi}-B_{p,s+i})=0 \:\:$ for all $i =1, \ldots , s$, and for all $p=1, \ldots , m$.
\end{description}
\end{th}

\mbox{}

\noindent {\bf Proof.} We denote by $L$ the Jacobian matrix of mapping (\ref{qpm}) with 
\begin{equation}
\label{lijdef}
	L_{ij} = \frac{\partial F_i}{\partial x_j} \equiv \partial_j F_i
\end{equation}
where
\begin{equation}
\label{fi}
	F_{i} = x_i \exp \left( \lambda _i + \sum _{j=1}^m A_{ij} 
	\prod _{k=1}^n x_k^{B_{jk}} \right) \equiv x_i \exp{(\varphi_i)} 
	\:\:\: , \:\: i = 1, \ldots , n=2s
\end{equation}
Making use of (\ref{lijdef}) and (\ref{fi}) we find:
\begin{equation}
\label{lijexp}
	L_{ij} = (\delta _{ij}+x_i \partial _j \varphi _i) \exp{( \varphi _i )} 
	\:\:\: , \:\: i,j = 1, \ldots , n
\end{equation}
We can then apply criterion (\ref{sympcond}--\ref{sympmat}) and find after some algebra that 
\[
	L^T \cdot S \cdot L = \left( \begin{array}{rr} 
		O_{s \times s} & -Q_{s \times s} \\
		Q_{s \times s} &  O_{s \times s} \end{array} \right)
\]
where
\begin{equation}
\label{quij}
	Q_{ij} = \sum_{k=1}^s \left( L_{s+k,s+i}L_{k,j} - L_{k,s+i}L_{s+k,j} \right)
	\:\:\: , \:\: i,j = 1, \ldots , s
\end{equation}
Therefore the symplectic condition now amounts to $Q_{ij}= \delta _{ij}$. Substituting 
(\ref{lijexp}) into (\ref{quij}) we arrive after some calculations to:
\[
	Q_{ij} = \delta _{ij} \exp{( \varphi _i + \varphi _{s+i})} +
		x_i \exp{( \varphi _i + \varphi _{s+i})} \partial _j \varphi _i +
		x_{s+j} \exp{( \varphi _j + \varphi _{s+j})} \partial _{s+i} \varphi _{s+j} + 
\]
\begin{equation}
\label{quij2}
		\sum_{k=1}^s x_k x_{s+k} 
		(\partial _{s+i} \varphi _{s+k}\partial _{j} \varphi _{k}-
		\partial _{s+i} \varphi _{k}\partial _{j} \varphi _{s+k})
		\exp{( \varphi _k + \varphi _{s+k})} \:\:\: , \:\: i,j = 1, \ldots , s
\end{equation}
Obviously it is necessary that $\: \varphi _i  + \varphi _{s+i} =$ constant for all $i=1, \ldots , 
s$. This implies immediately Condition (a) of the theorem. Consequently, we have $\varphi _i  + \varphi _{s+i} = \lambda _i + \lambda _{s+i}$. Substituting this result into (\ref{quij2}) the
following expression for $Q_{ij}$ is obtained:
\begin{equation}
\label{quij3}
	Q_{ij} = \delta _{ij} \exp{( \lambda _i + \lambda _{s+i})} +
		x_i \exp{( \lambda _i + \lambda _{s+i})} \partial _j \varphi _i +
		x_{s+j} \exp{( \lambda _j + \lambda _{s+j})} \partial _{s+i} \varphi _{s+j}
\end{equation}
Let us now consider, in particular, the case $i=j$. Then (\ref{quij3}) becomes:
\begin{equation}
\label{quiii}
	Q_{ii} = (1 + x_i \partial _i \varphi _i + x_{s+i} \partial _{s+i} \varphi _{s+i})
		\exp{( \lambda _i + \lambda _{s+i})}
\end{equation}
It is evident in (\ref{quiii}) that the only constant term multiplying the exponential is 1, and therefore this implies Condition (b) of the theorem. 

\noindent Taking Conditions (a) and (b) into account we arrive to:
\[
	Q_{ij} = \delta _{ij} +	x_i \partial _j \varphi _i + 
	x_{s+j} \partial _{s+i} \varphi _{s+j} \:\:\: , \:\: i,j = 1, \ldots , s
\]
Then the symplectic condition now becomes $x_i \partial _j \varphi _i + x_{s+j} \partial _{s+i} \varphi _{s+j}=0$ for all $i,j = 1, \ldots , s$. Substituting functions $\varphi _i$ according to their definition in (\ref{fi}) we arrive to:
\begin{equation}
\label{cond34}
	x_i \partial _j \varphi _i + x_{s+j} \partial _{s+i} \varphi _{s+j} = 
	\sum _{p=1}^m \left( \prod_{q=1}^n x_q^{B_{pq}} \right) 
	( A_{ip}B_{pj}x_ix_j^{-1} + A_{s+j,p}B_{p,s+i}x_{s+j}x_{s+i}^{-1})=0
\end{equation}
If we examine condition (\ref{cond34}) in the cases $i \neq j$ and $i=j$ we obtain Conditions (c) and (d) of the theorem, respectively. This completes the proof. \hfill {\Large $\Box$}

\mbox{}

Conditions (a-d) of Theorem \ref{th1} impose a very definite form on matrices $A$, $B$, $\lambda$ and $M$. This can be seen by means of some results that directly arise from such conditions: 

\begin{co}
\label{co1}
For every symplectic QP mapping the following properties hold: 
\begin{description}
\item[{\rm a)}]{\rm Rank}($B$)$\leq s$.

\item[{\rm b)}] {\rm Rank}($A$) $\leq$ {\rm Rank}($M$)$\leq s$.
\end{description}
\end{co}

Actually we can state:

\begin{co}
\label{co2}
A QP mapping (\ref{qpm}) is symplectic if and only if for every $p=1, \ldots , m$ the following conditions are satisfied:
\begin{description}
\item[{\rm a)}] Condition (b) of Theorem \ref{th1}.

\item[{\rm b)}] Row $p$ of $B$ has all entries equal to zero except two, which are $B_{p,i_p}$ and $B_{p,s+i_p}$, where index $i_p$ may change arbitrarily for different values of $p$ and
 $1 \leq i_p \leq s$.

\item[{\rm c)}] $B_{p,i_p}=B_{p,s+i_p}$.

\item[{\rm d)}] Column $p$ of $A$ has all entries equal to zero except two, given by $A_{i_p,p}$ and $A_{s+i_p,p}$.

\item[{\rm e)}] $A_{i_p,p}+A_{s+i_p,p}=0$.
\end{description}
\end{co}

The symplectic relations allow to demonstrate an interesting additional result:

\begin{pr}
\label{lm1}
For every symplectic QP mapping there exist $\,s\,$ conserved quantities $I_1, \ldots , I_s$ given by:
\begin{equation}
\label{ipes}
	I_i(x_1 , \ldots , x_n) = x_i (t) x_{s+i} (t) \:\: , \:\:\: i = 1 , \ldots , s
\end{equation}
\end{pr}

\mbox{}

\noindent {\bf Proof.} From conditions (a) and (b) of Theorem \ref{th1} it can be deduced that:
\[
	\ln \left( \frac{x_i(t+1)}{x_i(t)} \right) + 
	\ln \left( \frac{x_{s+i}(t+1)}{x_{s+i}(t)} \right) =0 \:\: , \:\:\: i = 1, \ldots , s
\]
This implies that $I_i=x_i(t)x_{s+i}(t)=x_i(0)x_{s+i}(0)$ and therefore is a constant 
quantity. \hfill {\Large $\Box$}

\mbox{}

The properties of the invariants associated to the rank degeneracy of matrix $M$ (such as those considered in Proposition \ref{lm1}) were generally analyzed in \cite{bl1}. The reader is referred to such reference for additional details.

In order to complement and clarify the exposition on the characterization of symplectic mappings it is convenient to present some brief examples. 

\mbox{}

\noindent {\bf Example 1.} As a first example consider the case $n=2$, with $m$ arbitrary. Let us apply Theorem 1. From Conditions (a) and (b) we find that matrix $M$ is of the form:
\[
	M = \left( \begin{array}{rrrr} 
		  \lambda_1 &  A_{11} & \ldots &  A_{1m} \\
		- \lambda_1 & -A_{11} & \ldots & -A_{1m} 
		\end{array} \right)
\]
It is then clear from the form of the QP equations for this mapping that the product 
$x_1(t)x_2(t)$ is conserved quantity, as anticipated by Proposition \ref{lm1}. Condition (c) of Theorem \ref{th1} does not apply, since here we have $i=j=1$ (note that $s=1$). From Condition (d) we obtain: 
\[
	A_{1p}(B_{p1}-B_{p2})=0 \:\: , \:\:\: p=1, \ldots ,m
\]
It is not possible to have any $A_{ip}=0$ because this implies a null column in matrix $A$. Therefore we must have $B_{p1}=B_{p2}$ for all $p$, namely:
\[
	B = \left( \begin{array}{cc} 
		B_{11} & B_{11} \\
		\vdots & \vdots \\
		B_{m1} & B_{m1}
		\end{array} \right)
\]
This is the general form of all the symplectic QP mappings in dimension 2.

\mbox{}

\noindent {\bf Example 2.} As a second example we may consider the case $n=4$ (or $s=2$) and 
$m=5$. One typical possibility allowed by the conditions of Theorem \ref{th1} (or equivalently Corollary \ref{co2}) is: 
\[
	M = \left( \begin{array}{cccccc}
			  \lambda_1 &     0   &    0    &     0   &  A_{14} &  A_{15}  \\
			  \lambda_2 &  A_{21} &  A_{22} &  A_{23} &  0      &  0       \\
			- \lambda_1 &     0   &    0    &     0   & -A_{14} & -A_{15}  \\
			- \lambda_2 & -A_{21} & -A_{22} & -A_{23} &  0      &  0     
		\end{array} \right)
\]
It is easy to verify from the form of the equations that there are two conserved products, as 
shown in Proposition \ref{lm1}:
\[
	I_1 = x_1(t)x_3(t) \:\:\: , \:\:\:\:\: I_2 = x_2(t)x_4(t)
\]
The application of the conditions of Theorem \ref{th1} (also the application of the rules given in Corollary \ref{co2}) is left to the reader. They lead to the following form of $B$:
\[
	B = \left( \begin{array}{cccc}
			0      &  B_{12} &    0   & B_{12}  \\
			0      &  B_{22} &    0   & B_{22}  \\
			0      &  B_{32} &    0   & B_{32}  \\
			B_{41} &     0   & B_{41} &    0    \\
			B_{51} &     0   & B_{51} &    0  
		\end{array} \right)
\]
Notice the relationship between the patterns of zeros for matrices $B$ and $A$, which is a characteristic feature of symplectic QP mappings.

\mbox{}

To conclude this section, it is interesting to present some additional results regarding the 
symplectic nature of QP mappings in the framework of the QP equivalence classes. The first one is the following:

\begin{co}
\label{co3}
For every QP mapping the class invariant $B \cdot M$ is the null $m \times (m+1)$ matrix.
\end{co}

Since the class invariant $B \cdot M$ is the $M$ matrix of the canonical Lotka-Volterra 
representative \cite{bl1} we also arrive to the following conclusion: 

\begin{co}
\label{co4}
The only symplectic mappings of Lotka-Volterra form are those having a null matrix $M$, namely the trivial mappings $x_i(t+1)=x_i(t)$, for $i=1, \ldots , n=2s$.
\end{co}

A simple counter-example allows the demonstration of our next corollary. For that, it suffices to consider the QP mappings of Example 1 and apply a QMT of matrix $C=$diag$(1,2)$. The 
conclusion is thus:

\begin{co}
\label{co5}
The property of being symplectic is not generally maintained in QP mappings after a QMT. In other words, it is not an invariant property in the QP equivalence classes.
\end{co}

On the other hand, it is certain with full generality that QMTs with matrices of the form $C= \mu I$, with $\mu \in I \!\! R - \{ 0 \}$, do preserve the symplectic property for QP mappings of arbitrary (even) dimension. This is verified by demonstrating that the conditions of Theorem \ref{th1} (or equivalently those of Corollary \ref{co2}) still hold after such transformations. The proof is straightforward and left to the reader. Therefore we obtain the last result of this section, which complements Corollary \ref{co5}: 

\begin{co}
\label{co6}
If a QP class of equivalence contains one symplectic mapping, then the class contains an infinity of symplectic mappings.
\end{co}

With the background provided by the results of this section, we can now focus on the issue of the solvability of QP symplectic mappings.

\mbox{}

\mbox{}

\begin{flushleft}
{\bf 4. Analytical solution of QP symplectic mappings}
\end{flushleft}

The aim of this section is twofold. First, the analysis of QP symplectic mappings will 
be completed by explicitly constructing their solutions with full generality. Second, this 
will be accomplished by means of the algebraic tools provided by the QP methodology, and therefore the construction of the solutions of QP symplectic mappings constitutes also a new application of the formalism. We thus arrive to the second main result of the paper:

\begin{th}
\label{th2} The explicit solution of every QP symplectic mapping is of the form:
\begin{equation}
\label{solut}
	\begin{array}{rcl}
		x_i(t)      & = & x_i(0)k_{i}^t \\
		x_{s+i}(t)  & = & x_{s+i}(0)k_{i}^{-t}
	\end{array}
\end{equation}
where $i=1, \ldots , s$ and $k_{i}>0$ for all $i$.
\end{th}

\mbox{}

\noindent {\bf Proof.} The proof is constructive, actually allowing the determination of the constants $k_{i}$ in (\ref{solut}). For arbitrary even $n$, let us consider a QP symplectic mapping of matrices $A$, $B$, $\lambda$ (and $M$) and consider also the following matrix:
\begin{equation}
\label{cker}
	C = \left( \begin{array}{rr} 
		I_{s \times s} & I_{s \times s} \\ O_{s \times s} & -I_{s \times s}
	    \end{array} \right)
\end{equation}
Notice that the last $s$ columns of $C$ actually constitute a basis of ker($B$). For the proof it is also useful to have in mind that $C=C^{-1}$. According to (\ref{mtqmt}) after the application of a QMT of matrix $C$ in (\ref{cker}) the result is a new QP mapping (not symplectic) of matrices:
\begin{equation}
\label{msol}
	M'=C^{-1} \cdot M = C \cdot M = 
	\left( \begin{array}{cccc}
		0 & 0 & \ldots  & 0 \\
		\vdots     & \vdots & \mbox{} & \vdots \\
		0 & 0 & \ldots  & 0 \\
		\lambda _1 & A_{11} & \ldots  & A_{1m} \\
		\vdots     & \vdots & \mbox{} & \vdots \\
		\lambda _s & A_{s1} & \ldots  & A_{sm} 
	\end{array} \right) 
\end{equation}
\begin{equation}
\label{bsol}
	B'=B \cdot C = 
	\left( \begin{array}{cccccc}
		B_{11} & \ldots  & B_{1s} &    0   & \ldots  &    0    \\
		\vdots & \mbox{} & \vdots & \vdots & \mbox{} & \vdots  \\
		B_{m1} & \ldots  & B_{ms} &    0   & \ldots  &    0 
	\end{array} \right) 
\end{equation}
Let us denote by $\{ y_1, \ldots , y_n \}$ the variables of the transformed QP system of matrices (\ref{msol}-\ref{bsol}). From the form of such matrices we can write the transformed system as follows:
\begin{eqnarray*}
		y_i(t+1)      & = & y_i(t) \\
		y_{s+i}(t+1)  & = & y_{s+i}(t) \exp \left( \lambda _i + \sum _{j=1}^m A_{ij} 
						\prod _{q=1}^s [y_q(0)]^{B_{jq}} \right) 
						\equiv k_{i} y_{s+i}(t)
\end{eqnarray*}
where $i=1, \ldots , s$ and we see that the $k_{i}$ are positive. The solution of this system is then 
\begin{equation}
\label{trsol}
	\begin{array}{rcl}
		y_i(t)      & = & y_i(0) \\
		y_{s+i}(t)  & = & y_{s+i}(0)k_{i}^{t}
	\end{array}
\end{equation}
Application to (\ref{trsol}) of the inverse QMT of matrix $C$ in (\ref{cker}) leads to the general solution (\ref{solut}) for the symplectic system. \hfill {\Large $\Box$}

\mbox{}

This completes the description of QP symplectic maps. It is worth checking that the existence of the $s$ conserved quantities (\ref{ipes}) now becomes apparent in the explicit solution 
(\ref{solut}). Notice also that the time behaviour of the variables appears correlated for every pair $\{ x_i,x_{s+i} \}$, for $i=1, \ldots ,s$, in such a way that only two possibilities exist, namely: {\em (i)} if $k_{i} \neq 1$ then one of the variables tends to zero while the other diverges; and {\em (ii)} if $k_{i} =1$ then both variables remain constant. Finally, one interesting consequence which is worth mentioning explicitly is the following: 

\begin{co}
\label{co7}
Symplectic QP mappings cannot present chaotic dynamics.
\end{co}

The solution procedure of Theorem \ref{th2} can be illustrated by means of the last example proposed in the previous section:

\mbox{}

\noindent {\bf Example 3:} Consider the QP symplectic system characterized in Example 2. We proceed to solve it. Note first that $\mbox{ker} (B) = \mbox{span} \{ (1,0,-1,0); 
(0,1,0,-1) \}$. Then we define $C$ as:
\begin{equation}
\label{ej3c}
	C = C^{-1} = \left( \begin{array}{ccrr} 1 & 0 & 1 & 0 \\ 0 & 1 & 0 & 1 \\ 
							0 & 0 & -1 & 0 \\ 0 & 0 & 0 & -1 
			\end{array} \right)
\end{equation}
If we perform the QMT of matrix $C$ the result is a QP mapping of matrices:
\[
	M'= \left( \begin{array}{cccccc}
			       0    &     0   &    0    &     0   &     0   &  0    \\
			       0    &     0   &    0    &     0   &     0   &  0    \\
			  \lambda_1 &     0   &    0    &     0   &  A_{14} &  A_{15}  \\
			  \lambda_2 &  A_{21} &  A_{22} &  A_{23} &  0      &  0     
		\end{array} \right) \:\:\: , \:\:\:\:
	B' = \left( \begin{array}{cccc}
			0      &  B_{12} & 0  & 0  \\
			0      &  B_{22} & 0  & 0  \\
			0      &  B_{32} & 0  & 0  \\
			B_{41} &     0   & 0  & 0    \\
			B_{51} &     0   & 0  & 0  
		\end{array} \right)
\]
Let $\{ y_1, y_2, y_3, y_4 \}$ be the variables of the transformed QP mapping. Then the mapping equations can be written as:
\begin{eqnarray*}
	y_1(t+1) & = & y_1(t) \\
	y_2(t+1) & = & y_2(t) \\
	y_3(t+1) & = & y_3(t) \exp \left( \lambda_1 + A_{14}(y_1(0))^{B_{41}} + 
		A_{15}(y_1(0))^{B_{51}} \right) \equiv k_1y_3(t) \\
	y_4(t+1) & = & y_4(t) \exp \left( \lambda_2 + A_{21}(y_2(0))^{B_{12}} + 
		A_{22}(y_2(0))^{B_{22}} + A_{23}(y_2(0))^{B_{32}} \right) \equiv k_2y_4(t)
\end{eqnarray*}
The solution of this mapping is: 
\begin{eqnarray*}
	y_1(t) & = & y_1(0) \\
	y_2(t) & = & y_2(0) \\
	y_3(t) & = & y_3(0) k_1^t \\
	y_4(t) & = & y_4(0) k_2^t
\end{eqnarray*}
Now let $\{ x_1, x_2, x_3, x_4 \}$ be the variables of the initial symplectic QP mapping. Making use of the inverse QMT (of matrix given by (\ref{ej3c})) we finally arrive to its solution:
\begin{eqnarray*}
	x_1(t) & = & x_1(0) k_1^t \\
	x_2(t) & = & x_2(0) k_2^t \\
	x_3(t) & = & x_3(0) k_1^{-t} \\
	x_4(t) & = & x_4(0) k_2^{-t}
\end{eqnarray*}
Notice that the invariants $I_1=x_1(t)x_3(t)$ and $I_2=x_2(t)x_4(t)$, already derived in Example 2, are now evident in the solution.

\mbox{}

We do not elaborate further on analytical results regarding the QP symplectic mappings. Instead, we proceed to conclude the work by presenting some final comments.

\mbox{}

\mbox{}

\begin{flushleft}
{\bf 5. Concluding remarks}
\end{flushleft}

We have seen that the symplectic case for QP discrete-time systems can be completely characterized and solved. This is to some extent remarkable, given that the presence of a common behaviour such as Hamiltonian chaos is therefore discarded in such family. On the other hand, we believe that the results presented provide an interesting illustration of the potentialities and flexibility of the QP methodology. In fact, this is to our knowledge the first time in the literature that it is possible to demonstrate a general result of this nature for a whole family of nonlinear mappings in arbitrary dimension $n$. Actually the results of this article also constitute a contribution not only from the point of view of the characterization of systems, but also for the characterization of dynamical behaviours associated to such systems. Of course, it is well-known that the presence of complex dynamical behaviours is ubiquitous in QP mappings when the Hamiltonian context is excluded. The 
non-symplectic case is certainly generic in the QP framework, and it is the most important from the point of view of many applications as well. Such situation is mostly unexplored at present, and the results just demonstrated also constitute a clear indication in the sense that future research on QP mappings must focus to a large extent on the generic non-symplectic possibilities.

In spite of the previous considerations, the field of symplectic QP mappings also offers relevant potentialities for future investigation. As an instance of these perspectives, it is worth recalling the close connection between QP differential systems and QP maps (which can actually be regarded as the discretized version of the former). The description of QP mappings as the discretization of QP differential systems was analyzed in detail in \cite{bl1} and it was already mentioned in the Introduction. The parameter-space characterization obtained in this paper allows the establishment of a close connection between symplectic QP maps and some families of QP differential systems (mainly characterized by a rank degeneracy in matrix $B$) which appear naturally in different problems related to integrability \cite{bvl1} and especially to normal form analysis \cite{bre2,slb2}. This is interesting, as far as two kinds of problems for which an analytical solution can be developed (series solutions for QP differential systems and (\ref{solut}) for QP symplectic maps) can be related. The analysis of the consequences of this parallelism and the possible transfer of results between both scenarios is just one instance of the open problems which may constitute the subject of future investigation.

\mbox{}

\mbox{}

\begin{flushleft}
{\bf Acknowledgements}
\end{flushleft}

This research has been supported by a Marie Curie Fellowship of the European Community programme {\em ``Improving Human Research Potential and the Socio-economic Knowledge Base'' \/} under contract number HPMFCT-2000-00421. B. H.-B. also acknowledges L. Brenig for his financial support and kind hospitality at the Physics Department of the ULB.

\pagebreak

\end{document}